# A Computational Model of the Electrically or Acoustically Evoked Compound Action Potential in Cochlear Implant Users with Residual Hearing

Daniel Kipping*, Yixuan Zhang, and Waldo Nogueira

*Abstract – Objective:* In cochlear implant users with residual acoustic hearing, compound action potentials (CAPs) can be evoked by acoustic (aCAP) or electric (eCAP) stimulation and recorded through the electrodes of the implant. We propose a novel computational model to simulate aCAPs and eCAPs in humans, considering the interaction between combined electric-acoustic stimulation that occurs in the auditory nerve. *Methods:* The model consists of three components: a 3D finite element method model of an implanted cochlea, a phenomenological single-neuron spiking model for electric-acoustic stimulation, and a physiological multi-compartment neuron model to simulate the individual nerve fiber contributions to the CAP. *Results:* The CAP morphologies closely resembled those known from humans. The spread of excitation derived from eCAPs by varying the recording electrode along the cochlear implant electrode array was consistent with published human data. The predicted CAP amplitude growth functions largely resembled human data, with deviations in absolute CAP amplitudes for acoustic stimulation. The model reproduced the suppression of eCAPs by simultaneously presented acoustic tone bursts for different masker frequencies and probe stimulation electrodes. *Conclusion:* The proposed model can simulate CAP responses to electric, acoustic, or combined electric-acoustic stimulation. It considers the dependence on stimulation and recording sites in the cochlea, as well as the interaction between electric and acoustic stimulation in the auditory nerve. *Significance:* The model enhances comprehension of CAPs and peripheral electric-acoustic interaction. It can be used in the future to investigate objective methods, such as hearing threshold assessment or estimation of neural health through aCAPs or eCAPs.

## I. INTRODUCTION

THE compound action potential (CAP) of the auditory nerve is a summation response from the auditory nerve fibers (ANFs) evoked by a brief stimulus. In cochlear implant (CI) users with residual acoustic hearing in the implanted ear (electric-acoustic stimulation; EAS), CAPs can be evoked both acoustically (aCAP) and electrically (eCAP) and be recorded through the electrodes of the CI. Previously, several studies have reported on the interaction between electric stimulation (ES) and acoustic stimulation (AS) in the same ear. These studies were conducted using electrocochleography [1], [2] or psychoacoustic experiments [3]–[7] in humans, as well as electrophysiological measures in animals [8]–[12]. They found that ES and AS can mask each other, resulting in reduced amplitudes of evoked responses and impaired perception of the stimuli, which may even limit the benefit in speech perception for EAS subjects [7]. Recently, another study reported significant reductions in eCAP responses when the electric probe was presented together with acoustic tone bursts [13], providing a unique opportunity to objectively explore the interaction between ES and AS at the level of the auditory nerve in human EAS subjects.

Computational models of the peripheral auditory system are widely used in auditory research today [14]–[17]. These models are especially useful in the context of electrophysiological responses, as they allow for the separation of interconnected effects [18]–[21] and a deeper understanding of the underlying physiology [22]–[29]. The aim of the present study is to develop a computational model of the CAP in response to ES and AS that accounts for electric-acoustic interaction in the auditory nerve.

Previous computational models of the CAP have been designed for either ES alone or AS alone and often assumed a "unitary response" for all ANFs [30]–[34]. The concept of a

This work is part of the project that received funding from the European Research Council (ERC) under the European Union's Horizon-ERC Program (Grant agreement READIHEAR No. 101044753 – PI: WN), was funded in part by the Deutsche Forschungsgemeinschaft (DFG, German Research Foundation) – SFB/TRR-298-SIIRI – Project-ID 426335750, and was supported by the DFG Cluster of Excellence EXC 2177/1 Hearing4all.

All authors are with the Hannover Medical School (MHH), Hannover, Germany, and with the Cluster of Excellence Hearing4All, Germany (*correspondence e-mail: kipping.daniel@mh-hannover.de).

unitary CAP response assumes that action potentials in all ANFs contribute with an identical "unitary" waveform pattern $U(t)$ to the recorded CAP. Under this assumption, the CAP response can be expressed as a convolution between the post-stimulus time histogram (PSTH) and $U$:

$$\text{CAP}(t) = \frac{1}{n_{\text{rep}}} \sum_j \left( \text{PSTH}_j * U \right)(t) \ . \tag{1}$$

In (1), $\text{PSTH}_j$ represents the spike times of ANF $j$ for $n_{\text{rep}}$ stimulus repetitions, which can be predicted from a spiking model of the ANF. The sum in (1) combines the contributions of all ANFs. In the context of the present study, (1) implies that the unitary response is postulated to be independent of the stimulation source, the morphology of the ANF in which the action potential occurred, as well as of the position of the recording electrode relative to that ANF. Various methods have been used to estimate $U$, such as measuring the PSTHs and the resulting CAP in animals and applying a deconvolution method to invert (1) [30], or parametrizing the shapes of both $U$ and the PSTH and fitting the parameters to CAP recordings [35], [36].

In reality, the assumptions for the unitary response are unlikely to be fulfilled. For example, eCAP amplitudes are largest when the recording site is adjacent to the stimulating site, and the amplitude decreases with larger separation of the two electrodes. This fact is commonly used for spread of excitation measurements with eCAPs [37]–[39]. Furthermore, it was hypothesized that CAP contributions for ES and AS may differ due to the locations along the ANFs where the action potentials are initiated. In the case of AS, the action potentials are elicited at the synapse between an inner hair cell (IHC) and the ANF, and propagate in central (orthodromic) direction along the entire length of the neuron. In contrast, for ES, the action potentials can be initiated at any point along the ANF, depending on the specific extracellular electric field produced by the stimulating electrode. As a result, action potentials for ES can split and propagate in both orthodromic and antidromic directions. For these reasons, we generalized (1) and replaced the unitary response with individual single-fiber CAP contributions (SFCCs) that were specific for each stimulation source $S$, recording electrode $R$, and ANF $j$:

$$\begin{aligned}\text{CAP}^{(S,R)}(t) = &\frac{1}{n_{\text{rep}}} \sum_j \left( \text{PSTH}_{\text{el},j} * \text{SFCC}_{\text{el},j}^{(S,R)} \right)(t) \\ &+ \frac{1}{n_{\text{rep}}} \sum_j \left( \text{PSTH}_{\text{ac},j} * \text{SFCC}_{\text{ac},j}^{(R)} \right)(t) \ .\end{aligned} \tag{2}$$

The two terms in (2) represent the contributions from ES and AS, respectively. A phenomenological spiking model for EAS [40] with a three-dimensional (3D) electrode-neuron interface based on a finite element method (FEM) model of the cochlea was used to predict the PSTHs for ES and AS. The SFCCs for ES and AS were obtained from a multi-compartment ANF model [41], [42]. This multi-compartment model was integrated into the 3D FEM model to simulate the excitation and propagation of action potentials along the ANFs, as well as the resulting potentials recorded on the CI electrodes.

## II. MODEL

The proposed model comprises a 3D voltage spread model based on FEM, a phenomenological neuron model for predicting the PSTHs in response to combined EAS, and a physiological multi-compartment neuron model for simulating the SFCCs.

This study involved the retrospective use of anonymized departmental patient data. According to the vote of the ethics committee of the Hannover Medical School (No. 1897-2013), no IRB approval is required for this type of use.

### A. Voltage Spread Model

A 3D FEM model of an average-sized human cochlea was used to simulate voltage transfer functions between the CI electrode contacts and 3000 ANFs. The FEM model was based on [25], [43]. The following compartments were distinguished in the FEM model: scala tympani, scala media, scala vestibuli, Reissner's membrane, basilar membrane, nerve tissue, and bone (Fig. 1a). An electric conductivity was assigned to each compartment of the FEM model (Fig. 1b). For the present study, the conductivities of the silicone CI array and the platinum CI electrode contacts were adapted to the values used in [44]. Furthermore, a cylindrical model of the auditory nerve trunk was attached at the base of the cochlea in the direction of the modiolar axis, extending the length of the auditory nerve by an additional 10 mm. A new model of a CI electrode array (CI24REH / Hybrid-L, Cochlear Ltd) tailored for EAS subjects was virtually inserted into the scala tympani. Pre- and post-op cone-beam computer tomography data of Hybrid-L users available at Hannover Medical School (MHH) were analyzed to construct a typical insertion path of the electrode array with the average insertion depth of 14.7 mm. The electrode array's 22 half-ring contacts were distributed over an active length of 14.5 mm according to the specifications of the manufacturer, facing the modiolus.

The cochlea model was placed in the center of a solid bone sphere with a diameter of 35 mm. The surface of the sphere was used as the ground for electrostatic FEM simulations, emulating a reference electrode located at an infinitely distant location from the cochlea. Electrostatic FEM simulations were performed with COMSOL Multiphysics 5.6 (COMSOL AB, Stockholm, Sweden) to predict the voltage distribution $V_0^E(x, y, z)$ in the 3D model when a fixed current of $I_0 = 1 \, \mu A$ was delivered to the CI electrode $E$.

The electrical conductivity of the solid bone sphere was fitted based on transimpedance matrices, a clinical measure used to estimate the intra-cochlear voltage spread [25]. Simulated transimpedance matrices were compared to clinical transimpedance matrices of 5 Hybrid-L users at MHH for a range of bone conductivities. The bone tissue's electric conductivity was optimized to minimize the root mean square error between simulated and measured transimpedance matrices, resulting in a fitted bone conductivity of 0.0116 S/m (Fig. 1b).

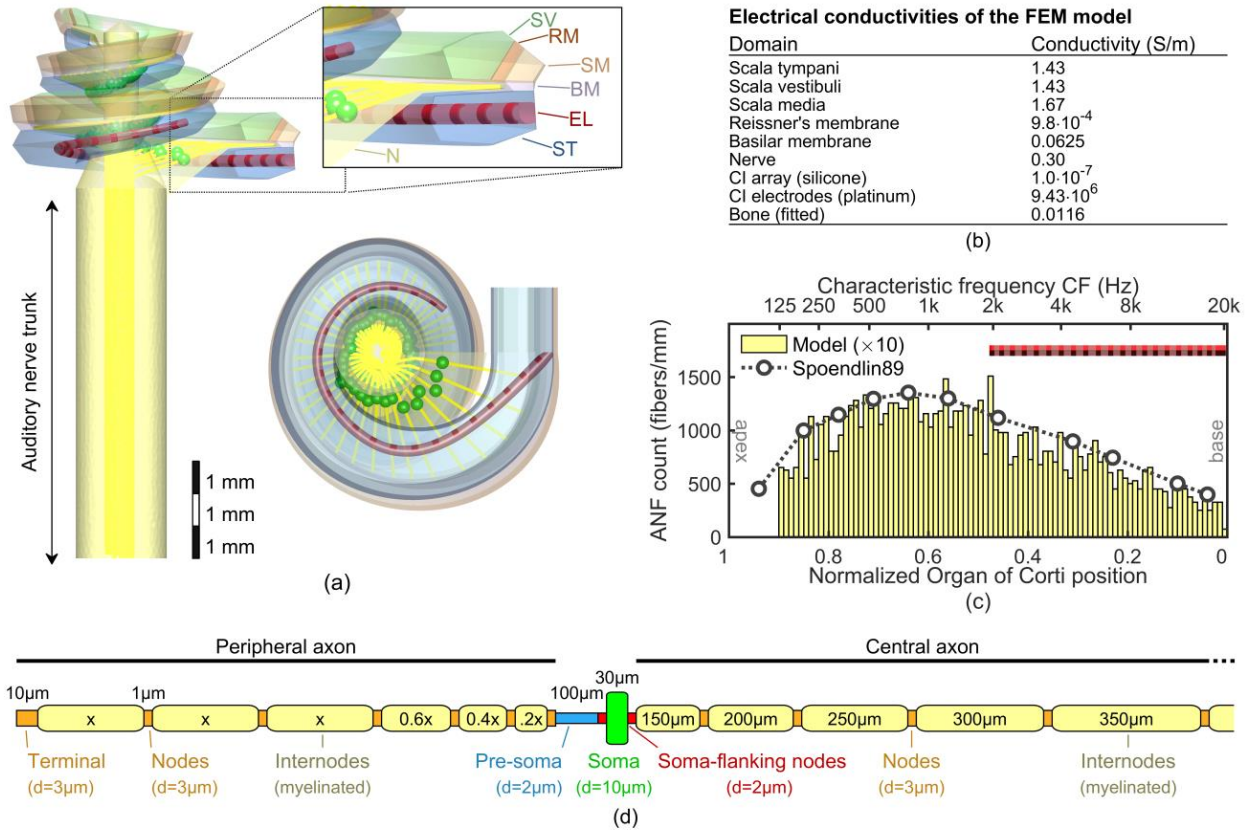

Fig. 1. 3D finite element method (FEM) model and auditory nerve fiber (ANF) morphology. **(a)** 3D FEM model used to simulate the voltage spread in the cochlea and auditory nerve. 3000 ANFs were placed in the 3D model (yellow lines), with the somata (green) located in Rosenthal's canal. The FEM model was composed of the scala vestibuli (SV), Reissner's membrane (RM), scala media (SM), basilar membrane (BM), CI electrode array and electrode contacts (EL), scala tympani (ST), and nerve tissue (N) and was embedded in a bony sphere (d=35 mm). **(b)** Electrical conductivities assigned to each of the compartments. The conductivity of the surrounding bone sphere was optimized based on simulated and measured transimpedances. **(c)** Density of ANFs along the length of the organ of Corti (1=base, 0=apex) for the model compared to data reported by [46]. For the visualization, ANF counts for the model were multiplied by a factor of 10 since each modeled ANF represented 10 ANFs in a real cochlea. Also shown is the location of the CI electrode contacts (red), with the apical electrode corresponding to 1975 Hz. **(d)** ANF morphology of [42]. Each ANF consisted of a peripheral axon (with terminal, myelinated internodes, and unmyelinated nodes), an unmyelinated pre-somatic region, the unmyelinated soma flanked by two thin nodes, and the central axon (with nodes and internodes). The base length of the peripheral internodes ("x") was scaled to fit the distance between the Organ of Corti and the soma as proposed by [42]. Additional nodes and internodes (350 µm) were added to the central axon to fit the distance between the soma and the bottom of the auditory nerve trunk.

## B. 3D Nerve Fiber morphology

The pathways of 3000 myelinated Type I ANFs were embedded within the nerve tissue of the FEM model (Fig. 1a). Each modeled ANF represented 10 ANFs in a real cochlea, totaling 30000 ANFs. The density of ANFs was varied along the length of the organ of Corti (OC) according to data from [45], [46] (Fig. 1c). The density of ANFs was highest in the first half of the second cochlear turn, corresponding to around 55%–70% of the OC length measured from base to apex. Each ANF was assigned a characteristic frequency based on the tonotopic map of Greenwood [47]. The range of characteristic frequencies was clipped at 125 Hz to 20 kHz to match the frequency range of the phenomenological spiking model [40], [48]. The ANFs first followed the Rosenthal's canal radially from their position at the OC towards the modiolus, and then followed the modiolus towards the internal auditory canal and auditory nerve trunk (Fig. 1a). Each ANF pathway was defined as a set of points representing compartments of the neuron, following the

ANF morphology proposed by [42] (Fig. 1d). It consisted of a peripheral terminal connected to the OC, at least six unmyelinated Nodes of Ranvier ("nodes"), a pre-somatic compartment, the soma located in Rosenthal's canal, and at least 38 nodes in the central axon. As in [42], the number of nodes in the peripheral and central axons as well as the base length "x" between adjacent nodes in the peripheral axon were based on the length of the processes. The internodes separating the nodes were myelinated and considered to be perfectly insulating.

The 3D voltage distributions $V_0^E(x, y, z)$ were sampled at the compartment centers $i$ of the ANFs $j$, resulting in 2D voltage matrices $V_{0,ij}^E$. Negligibility of capacitive effects was assumed in all materials of the 3D FEM model. Therefore, the voltage distribution resulting from an arbitrary time-varying stimulation current $I(t)$ could be derived by linear scaling:

$$V_{ij}^E(t) = I(t) \cdot V_{0,ij}^E / I_0. \quad (3)$$

### C. Simulation of single fiber CAP contributions

The SFCCs were simulated with the nonlinear Schwarz & Eikhof-Frijns multi-compartment model (SEF model; unmyelinated cell body condition; see [41]) coupled to the 3D FEM model as proposed by [23]. The equations for the SEF model can be found in [41], and a comprehensive review and an implementation in Python are provided in [49]. In summary, for a given stimulus the SEF model can simulate the time course of the cross-membrane potential and the cross-membrane currents in all compartments of the ANF. Separate SFCCs were simulated for each ANF $j$, stimulation source $S$, and recording electrode $R$. For ES via electrode $E$, the input to the SEF model was the extracellular voltage distribution (3), where the stimulation current $I(t)$ was a cathodic-leading biphasic pulse with a duration of 25 µs per phase and an inter-phase gap of 7 µs. For AS, the input was an excitatory post-synaptic current induced intracellularly into the peripheral terminal compartment. For simplicity, a rectangular current injection with the same duration of 25 µs was used. The stimulation levels for ES and AS were adjusted iteratively until an action potential was generated and conducted to the ANF's central end (supra-threshold condition). The model output was the total cross-membrane current $I_{\text{mem},i}^S(t)$ in each compartment $i$.

The cross-membrane currents served as current sources to predict the corresponding voltage recorded through electrode $R$ by means of the reciprocal theorem:

$$v_{\text{rec},ij}^{S,R}(t) = I_{\text{mem},i}^S(t) \cdot V_{0,ij}^R / I_0 . \qquad (4)$$

Here, the resistance matrix $V_{0,ij}^R / I_0$ that represents the voltage predicted at compartment $i$ of ANF $j$ when stimulating electrode $R$ was used to describe the reciprocal current path from the ANF compartment to the recording electrode. These contributions were summed over all compartments to obtain the supra-threshold recording from ANF $j$:

$$V_{\text{rec}}^{S,R}(t) = \sum_i v_{\text{rec},ij}^{S,R}(t) . \qquad (5)$$

The potentials in (5) still contain two types of artifacts: the stimulation artifact and a "border" artifact that occurs when the action potential reaches the ANF's central end. To remove the stimulation artifact, sub-threshold artifact scaling was used [23]. This method eliminates the stimulation artifact by taking advantage of the fact that it scales linearly with the stimulation level. For this purpose, each potential waveform (5) was predicted at a supra-threshold level $I_{\text{supra}}$ containing stimulation artifact as well as the neuronal response, and at a sub-threshold level $I_{\text{sub}}$ containing only the artifact. The SFCC was obtained by subtracting the scaled sub-threshold artifact from the supra-threshold response:

$$\text{SFCC}_j^{S,R}(t) = V_{\text{rec(supra)},j}^{S,R}(t) - I_{\text{supra}}/I_{\text{sub}} \cdot V_{\text{rec(sub)},j}^{S,R}(t). \quad (6)$$

When compared to commonly used artifact elimination approaches such as the masker-probe paradigm [50], sub-threshold artifact scaling has the advantage that the stimulation artifact is already removed at the stage of the SFCC. This allows to analyze artifact-free SFCCs, and subsequently requires only a single simulation run for the probe PSTH.

To remove the border artifact, the simulated SFCC waveform (6) was replaced with a cubic spline in a short 250 µs time interval around the time when the action potential arrived at the ANF's central end. More details and an illustration of the artifact removal can be found in the Supplementary Material (Suppl. Fig. 1).

### D. Simulation of spike times (PSTHs)

PSTHs were predicted with a phenomenological single-ANF spiking model for EAS [40] (Fig. 2a). The model takes an electric and an acoustic stimulus as input and produces stochastic spike times for a specific ANF. It consists of two sub-models: a phenomenological model of the auditory periphery for AS [48], coupled to a double integrate-and-fire point neuron model for ES [51]. The electric and the acoustic spiking submodels interact through the refractoriness of the ANF. The "alternative coupling" method described in [40] was used, because with this setting the model predicted two separate PSTHs for the spikes evoked by ES or AS (see eq. (2) and Fig. 2a). The electric submodel predicts spike times for direct electroneural excitation of the ANF but does not consider electrophonic stimulation of remaining hair cells [52]. Since electrophony is improbable in human EAS users due to their severe high-frequency hearing loss [6], [53], the model likely represents the stimulation mode relevant for human EAS subjects.

For the present study, a 3D electrode-neuron interface for ES corresponding with the 3D FEM model was added to the electric spiking model as suggested in [54]. The 3D FEM model was used to derive scaling factors $s_j^E$ which represented the amount of current effectively induced in each individual ANF $j$ by stimulating electrode $E$ (Fig. 2b). The scaling factors were derived from the activating function [55],

$$\begin{aligned} \text{AF}_{ij}^E = &\ G_{(i-1)j} \cdot \left( V_{0,(i-1)j}^E - V_{0,ij}^E \right) / C_{ij} \\ &- G_{ij} \cdot \left( V_{0,ij}^E - V_{0,(i+1)j}^E \right) / C_{ij} , \end{aligned} \qquad (7)$$

where the axonal conductance $G_{ij}$ between compartments $i$ and $(i+1)$ and the membrane capacitance $C_{ij}$ of compartment $i$ were taken from the ANF morphology of [42]. Equation (7) represents the amount of current induced in each compartment $i$ of ANF $j$ by the extracellular voltage $V_0^E$. As in [54], it was hypothesized that the excitation of an ANF $j$ was most likely to occur in the compartment $i_{\max}$ where the largest amount of current is induced, i.e. where (7) has the largest absolute value:

$$i_{\max}(j) = \text{argmax}_i \left( \left| \text{AF}_{ij}^E \right| \right) . \qquad (8)$$

The argmax function in (8) considered all compartments in the peripheral and central axons, but excluded the pre-somatic and somatic compartments as well as the nodes directly adjacent to the soma (compare Fig. 1d). The reason was that these compartments exhibited abnormally high values of $\left| \text{AF}_{ij}^E \right|$ due to sudden changes in the axonal conductance $G_{ij}$. Finally, the

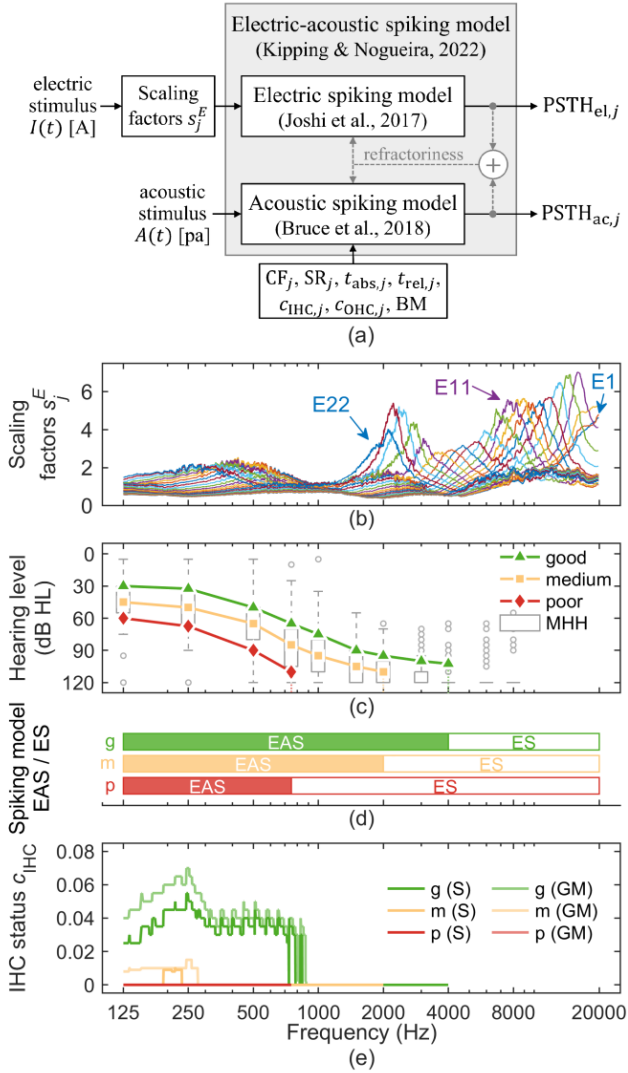

Fig. 2. Overview of the single-fiber spiking model used to predict spike times in response to electric-acoustic stimulation. **(a)** Block diagram of the spiking model for auditory nerve fiber (ANF) $j$ as proposed by [40]. The inputs are the stimulus waveforms for electric and acoustic stimulation, and the outputs are two post-stimulus time histograms (PSTHs) for the electrically and the acoustically evoked spikes. Electric and acoustic stimulation interact through the ANF's refractoriness. Parameters for the acoustic spiking model are characteristic frequency (CF), spontaneous discharge rate (SR), absolute and relative refractory periods ($t_{abs}$ and $t_{rel}$), inner and outer hair cell status ($c_{IHC}$ and $c_{OHC}$), and basilar membrane tuning (BM). A 3D electrode-nerve interface was added to the electric spiking model in terms of the scaling factors $s_j^E$. **(b)** Absolute value of the scaling factors $s_j^E$ that were applied to the electric stimulus. Different colors indicate different stimulating electrodes $E$. Peaks correspond with ANFs that were most sensitive to stimulation of the corresponding electrode. **(c)** Clinical audiograms of $N = 110$ EAS users from Hannover Medical School (MHH) (boxplot) and three hearing loss conditions defined for the acoustic spiking model. **(d)** Model variants used to predict PSTHs for each ANF. The electric-acoustic spiking model shown in (a) was used for ANFs with hearing levels below 120 dB HL ("EAS"). ANFs outside the range of residual hearing were simulated with the electric spiking model without the additional acoustic component ("ES"). g – good, m – medium, p – poor residual hearing. **(e)** Inner hair cell status $c_{IHC}$ as derived from the audiograms in (c) when the basilar membrane tuning was based on data of Shera et al. ("S"; [56]) or Glasberg and Moore ("GM"; [57]). The outer hair cell status $c_{OHC}$ was zero across the whole frequency range for all conditions (not shown).

scaling factors were computed as

$$s_j^E = \alpha \cdot \text{AF}_{i_{\max}(j), j}^E \,, \tag{9}$$

and could be both positive or negative, representing the two directions of current flow in the ANF. The global normalization factor $\alpha$ in (9),

$$\alpha = \frac{1}{N_{anf}} \frac{1}{N_E} \sum_j \sum_E \left| \text{AF}_{i_{\max}(j), j}^E \right|^{-1} \,, \tag{10}$$

with $N_{anf} = 3000$ and $N_E = 22$, ensured that the average input dynamic range of the electric spiking model (e.g. threshold and relative spread of each ANF) was preserved. The resulting scaling factors for all electrodes are shown in Fig. 2b.

To simulate the spiking of ANF $j$ in response to a current $I(t)$ delivered through electrode $E$, the input stimulus of the electric spiking model was

$$I_j^E(t) = s_j^E \cdot I(t) \,. \tag{11}$$

Equation (11) naturally implemented a variation in electric single-ANF thresholds and dynamic ranges. For this reason, the fixed membrane capacitances from the original electric spiking model [51] were applied instead of the randomized membrane capacitances proposed in [40].

According to [40], [51], the spike latencies predicted by the electric spiking model for single pulse stimuli were about 200 μs too low. To balance the latencies produced by the acoustic and the electric spiking models, a constant offset of 200 μs was added to the spike times predicted by the electric spiking model.

To account for the typical sloping hearing loss of EAS subjects in the acoustic spiking model, clinical audiogram data from 110 Hybrid-L EAS users at MHH were analyzed. Three representative audiograms were defined for the model: "good" residual hearing (20th percentile of the data), "medium" residual hearing (50th percentile), and "poor" residual hearing (80th percentile; Fig. 2c). It was assumed that IHCs and outer hair cells (OHCs) were only present in regions with residual hearing (hearing level <120 dB HL). Therefore, the PSTHs for ANFs with audiometric thresholds below 120 dB HL were simulated with the combined electric-acoustic spiking model as shown in Fig. 2a. Conversely, the PSTHs for the high-frequency ANFs with hearing loss ≥120 dB HL were simulated with the electric spiking model alone, without the acoustic component [40]. The highest tonotopic frequencies with surviving hair cells were 4000 Hz, 2000 Hz, and 750 Hz for good, medium, and poor residual hearing (Fig. 2d).

The parameters for IHC and OHC status ($c_{IHC}$ and $c_{OHC}$) were derived from the audiograms in Fig. 2c using the MATLAB function fitaudiogram2 from [48]. A value of 1 corresponded to normal hair cell function, whereas a value of 0 indicated maximal impairment. In the acoustic spiking model, the tuning of the basilar membrane (BM) can be based on data of either Shera et al. ("S"; [56]) or Glasberg and Moore ("GM"; [57]). Combined with the three hearing loss conditions good, medium, and poor, this resulted in a total of six different sets of IHC and OHC status parameters (Fig. 2e for $c_{IHC}$). For all conditions, the audiograms resulted in an OHC status $c_{OHC}$ of zero across the whole frequency range, indicating a complete loss of OHC function that leads to broadened tuning curves and loss of compression and suppression for AS [58].

The remaining parameters of the spiking model – spontaneous spike rate $SR_j$ and absolute and relative refractory periods $t_{abs,j}$ and $t_{rel,j}$ – were assigned from the same random distributions as in [40].

### E. Construction of CAPs

CAPs were constructed based on the PSTHs from the electric-acoustic spiking model and the SFCCs for ES and AS as stated in (2). Additionally, the results of (2) were multiplied by a factor of 10 to compensate for each modeled ANF representing 10 real ANFs in a physiological cochlea. For ANFs with CFs outside the range of residual hearing (see Fig. 2d, "ES"), only the first term in (2) was considered since no acoustic spiking model was used for these ANFs.

### F. Model availability

The described modeling framework was implemented in MATLAB and is available at GitLab[1] and Zenodo[2]. The repository contains the data and code necessary to perform the simulations presented in this paper, as well as to perform own simulations with custom stimuli.

## III. METHODS

### A. Estimation of threshold and comfortable loudness levels

The thresholds (T-level) and most comfortable loudness levels (M-level) were estimated for both ES and AS to set appropriate levels for the input stimuli. For ES, the T-level and the M-level were defined as the current levels where the excited ANFs covered a cumulative length of 1 mm or 4 mm along the OC, respectively [42], [59]. For this purpose, an ANF was considered as "excited" when the stimulation level was above its single-ANF threshold, which is the current level at which an individual ANF responds with a spike in 50% of the cases. The T-levels for AS were defined in the same way as for electric stimuli. For acoustic M-levels, the criterion of 4 mm excited OC length was found to be unreliable, as it would result in unrealistically low acoustic dynamic ranges (DRs). Therefore, M-levels for AS were estimated using experimental data that was available from previous studies with EAS subjects conducted in our lab [4]–[7], [60]. The data included 121 audiometric pure tone T-levels at different frequencies in dB HL and the corresponding acoustic DRs. The DR was defined as the range from the T-level to the M-level in dB. The data was pooled across subjects and stimulation frequencies. A linear fit of the acoustic DR vs. T-level data resulted in

$$DR(T_{dB\,HL}) = 51.4 - 0.4 \cdot T_{dB\,HL}. \quad (12)$$

The acoustic T-levels $T_{1mm}$ from the model defined via the

1 mm criterion were converted from dB SPL to dB HL by subtracting the threshold $T_{NH,1mm}$ obtained from the model with a "normal hearing" cochlea condition without hearing loss as a reference: $T_{dB\,HL} = T_{1mm} - T_{NH,1mm}$. For each hearing loss condition and acoustic stimulus, the acoustic M-level was then estimated as

$$M(T_{1mm}) = T_{1mm} + DR(T_{dB\,HL}), \quad (13)$$

where $DR(\cdot)$ is the linear fit (12).

### B. Experiments

Three experiments were conducted using the proposed CAP model: ES alone (experiment 1), AS alone (experiment 2), and acoustic masking of ES (experiment 3).

In experiment 1, ES consisted of single biphasic cathodic-leading pulses with a phase duration of 25 µs and an inter-phase-gap of 7 µs. eCAP waveforms were predicted for all electrodes, and eCAP amplitudes were measured from the first negative peak (N1) to the following positive peak (P1). The recording electrode $R$ was positioned two contacts apical from the stimulating electrode $S$, except for the two most apical electrodes, where $R$ was shifted by two contacts in the basal direction. eCAP amplitude growth functions (AGFs) were obtained by adjusting the stimulation levels from below T-level to M-level. The simulated eCAP AGFs were compared with clinical data from MHH and experimental data from [61]. Additionally, eCAPs were analyzed for different recording electrodes, and the resulting spatial spread profiles were compared to data of [37], [38].

Experiment 2 included AS with 100 µs clicks, 6.3 ms chirps (with the frequency rising from 450 Hz to 10 kHz [62], [63]), and 20 ms tone bursts at 500 Hz with 2 ms $cos^2$ ramps. To account for the frequency response of the earphone, all stimuli presented through AS were recorded from the output of an ER3 insert earphone (Intelligent Hearing Systems, Miami, FL) connected to a Phone-Amp G103 headphone amplifier (Lake People electronics GmbH, Konstanz, Germany) and an external NI-DAQ sound card (National Instruments, Austin, TX) before presenting them to the model. The recordings were made using an ER-7C probe microphone system (Etymotic Research, Inc., Elk Grove Village, IL) linked to a PicoScope 5443A oscilloscope (Pico Technology Ltd, Cambridgeshire, UK). In the simulations, AS was presented both with rarefaction and condensation polarities, and the average CAP waveform for both polarities was analyzed. aCAP AGFs were derived by varying the stimulation level from 45 dBnHL to 85 dBnHL (where the dBnHL scale denotes dB relative to the T-level obtained with a "normal hearing" cochlea model where no hearing loss was applied). Recording was conducted through the most apical electrode, E22. The predictions of the model were compared to experimental data of [63]. To be consistent with the post-processing of [63], all aCAP waveforms generated for experiment 2 were filtered using a second-order low pass Butterworth filter at a cut-off frequency of 500 Hz.

## Experiment 1: electric stimulation

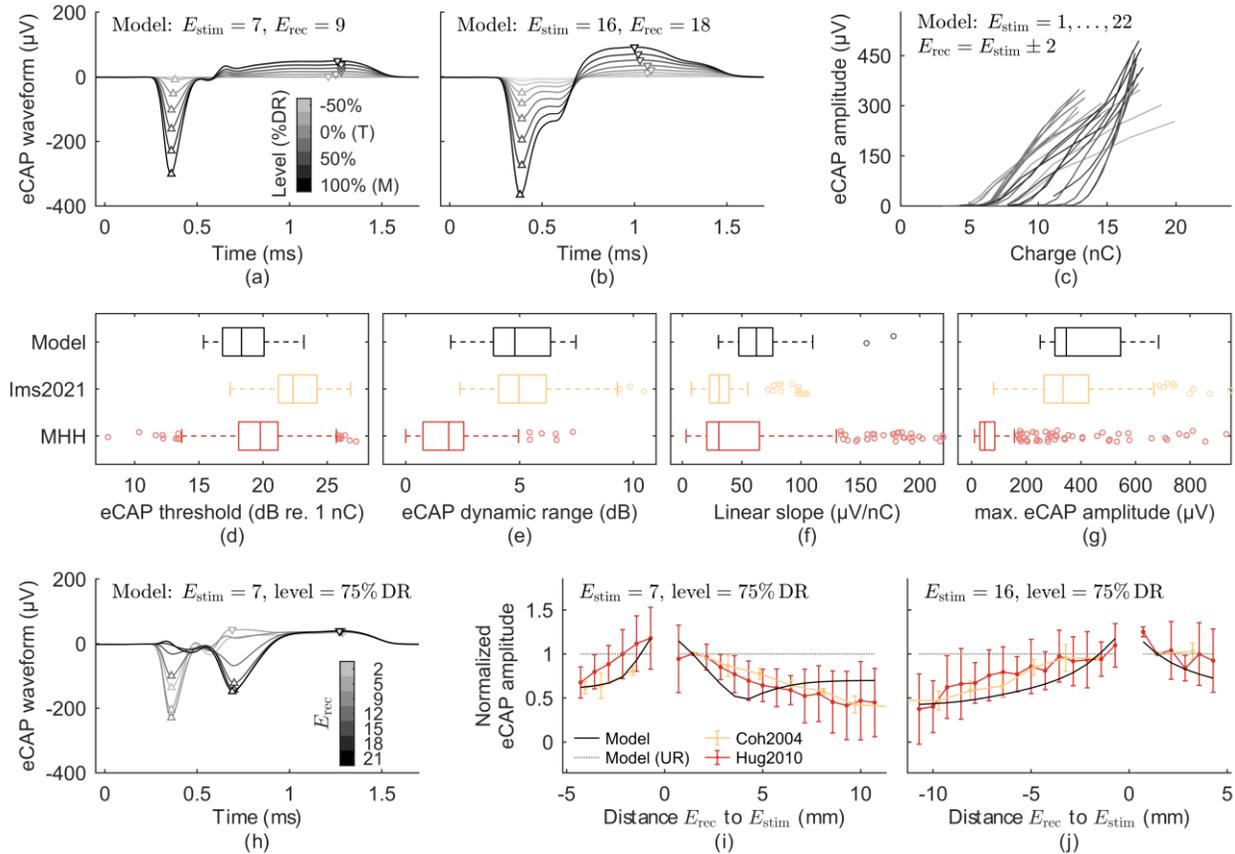

Fig. 3. Results for experiment 1 on the electrically evoked compound action potential (eCAP). **(a),(b)** eCAP waveforms predicted for stimulation at a basal (a) and at an apical (b) electrode and current levels between $-50\%$ DR and $100\%$ DR. N1 and P1 peaks are denoted with triangle markers. **(c)** eCAP amplitude growth functions (AGFs) predicted for all electrodes when $E_{rec}$ was 2 contacts apical to $E_{stim}$ (or 2 contacts basal to $E_{stim}$ for electrodes 21 and 22). The range of current levels was identical to (a) and (b). **(d)–(g)** Analytical measures (eCAP thresholds, dynamic range, linear slope, and maximum amplitude) derived by fitting the eCAP AGFs with sigmoid functions. Model predictions were compared to experimental data of Imsiecke et al. ("Ims2021", $N = 142$; [61]) and clinical data of EAS users at Hannover Medical School ("MHH", $N = 453$). Limitations regarding the recording procedures of the human data are discussed in the main text. **(h)** Example of the effect of changing the recording electrode on predicted eCAP waveforms when stimulation was fixed at E7 at a level corresponding with 75% DR. **(i),(j)** eCAP spatial spread for basal (i) and apical (j) stimulation at 75% DR. Model predictions ("Model") were compared to experimental data of Cohen et al. ("Coh2004"; [37]) and Hughes et al. ("Hug2010"; [38]). Error bars depict the mean and standard deviation of the data. Models based on the unitary response approach are not sensitive to changes in the recording site ("Model (UR)"). The results in this figure were obtained for the condition with medium residual hearing and BM tuning of Shera et al. ("S").

In experiment 3, the electric single pulse from experiment 1 was used as a probe stimulus, and 20 ms acoustic tone bursts as employed in experiment 2 were used as masker stimuli. The tone bursts were presented at 125 Hz, 250 Hz, 500 Hz, 1000 Hz, and 2000 Hz and recorded from an insert earphone with the same setup as for experiment 2. Acoustic maskers in the model were presented with rarefaction and condensation polarities, and the average CAP waveform for both polarities was analyzed. The electric probe was presented through the three apical electrodes E22, E21, and E20, and the CAP was recorded through electrodes E21, E22, and E21, respectively. As in the study of [13], the electric probe was separated from the acoustic masker onset by a 5 ms masker-to-probe interval, and both stimuli were presented at M-level. CAPs were obtained in three conditions: unmasked (ES), masked (AS+ES), and masker-only (AS). A "derived electric" response was calculated by subtracting the masker-only response from the masked response: derived ES = (AS + ES) − AS. The extent of acoustic masking was measured in terms of the reduction in CAP amplitude of the derived electric response compared to the unmasked electric (ES) response. The model predictions were compared to the results presented in [13].

## IV. RESULTS

### A. Single-fiber CAP contributions

The shape and amplitude of the predicted SFCCs differed significantly between ANFs and was dependent on the choice of the recording electrode as well as, in case of ES, the stimulating electrode. These effects are illustrated in the Supplementary Material (Suppl. Fig. 2).

### B. Experiment 1: eCAP AGFs

Experiment 1 investigated eCAP responses to single biphasic

## Experiment 2: acoustic stimulation

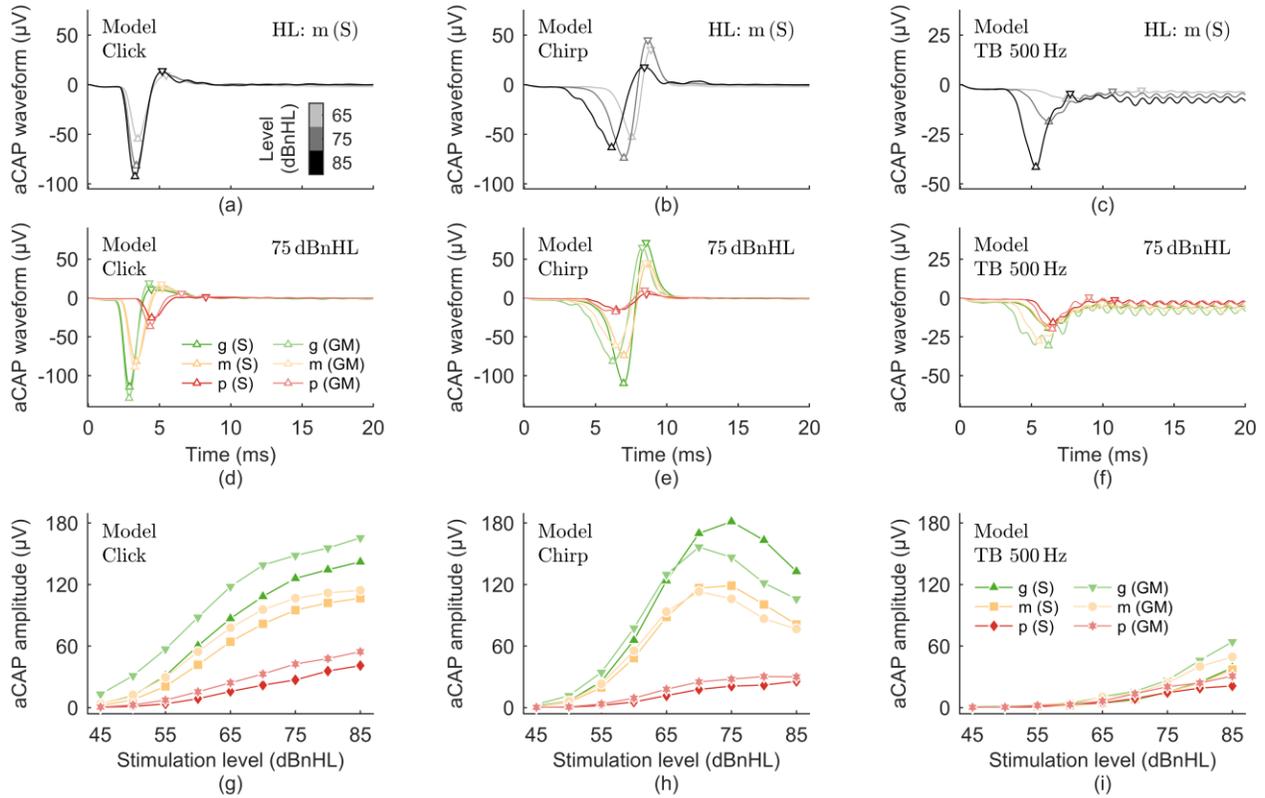

Fig. 4. Results for experiment 2 on the acoustically evoked compound action potential (aCAP). Stimuli were clicks (left column), chirps (middle column), and 500 Hz tone bursts (right column) and the recording electrode was E22. **(a)–(c)** Predicted aCAP waveforms for different stimulation levels in the condition with medium residual hearing and BM tuning of Shera et al. ("m (S)"). **(d)–(f)** Predicted aCAP waveforms at 75 dBnHL for all hearing loss conditions (g – good, m – medium, p – poor; BM tuning according to Shera et al. (S) or Glasberg & Moore (GM)). **(g)–(i)** Predicted aCAP amplitude growth functions for all hearing loss conditions. Compare Fig. 11A–C in [63] for the human data.

pulses (Fig. 3). The simulated eCAP waveforms closely resembled the physiological shape of the response, exhibiting a first negative N1 peak followed by a positive P1 peak (Fig. 3 a,b). The predicted eCAP amplitudes increased as the stimulation level was raised. eCAP AGFs were simulated for all 22 electrodes, positioning the recording electrode 2 contacts apical to the stimulating electrode (or 2 contacts in the basal direction for electrodes 21 and 22). The current levels were expressed in terms of the applied charge (Figure 3c).

Sigmoidal fits (14) with fitting parameters $A$, $B$, and $C$ were employed to characterize the eCAP AGFs:

$$V(q) = \frac{A}{1 + \exp(-(q - B)/C)} \ . \tag{14}$$

The fits were used to quantify the eCAP threshold ($B - 2C$), dynamic range (from threshold to $B + 2C$ in dB), linear slope ($A/(4C)$), and maximum amplitude ($A$) [61]. These characteristics are compared to two data sets of human eCAP AGFs (Fig. 3d–g). The first data set included 142 eCAP AGFs from 16 EAS users with MED-EL Flex 28, 24, 20, or 16 implants ("Ims2021"; [61]). This data set utilized a blanking technique to reduce stimulus artifacts in the recordings, potentially resulting in an underestimation of the absolute eCAP amplitude $A$ (Fig. 3g) and the linear slope (Fig. 3f) [61]. The second data set consisted of 453 clinical eCAP AGFs from 54 EAS subjects at MHH who were implanted with a Hybrid-L CI electrode array ("MHH"). These recordings utilized a masker-probe paradigm [64] to eliminate the stimulus artifact. This paradigm, according to the clinical NRT protocol on Cochlear devices, involves presenting a masker pulse 10 current units (about 1.6 dB) above the probe level, which restricts the usable current range of the probe stimulus. As a result, there is potential for underestimation of eCAP dynamic range (Fig. 3e), linear slope (Fig. 3f), and maximum amplitude (Fig. 3g). Given the uncertainties in the different recording procedures, the predicted eCAP AGF characteristics generally aligned with the two datasets.

To investigate the spatial spread of the eCAP responses, a fixed electrode was stimulated at a level of 75% DR and the recording site was varied along the electrode array. Altering the recording electrode significantly affected the morphology of the predicted eCAP waveforms (Fig. 3h). Following the protocol outlined in [38], the eCAP amplitudes were normalized to the amplitude recorded when the recording site was 2 contacts apical to the stimulation site. Fig. 3i,j compare these normalized spatial spread profiles for a basal and an apical electrodes with

## Experiment 3: electric-acoustic stimulation

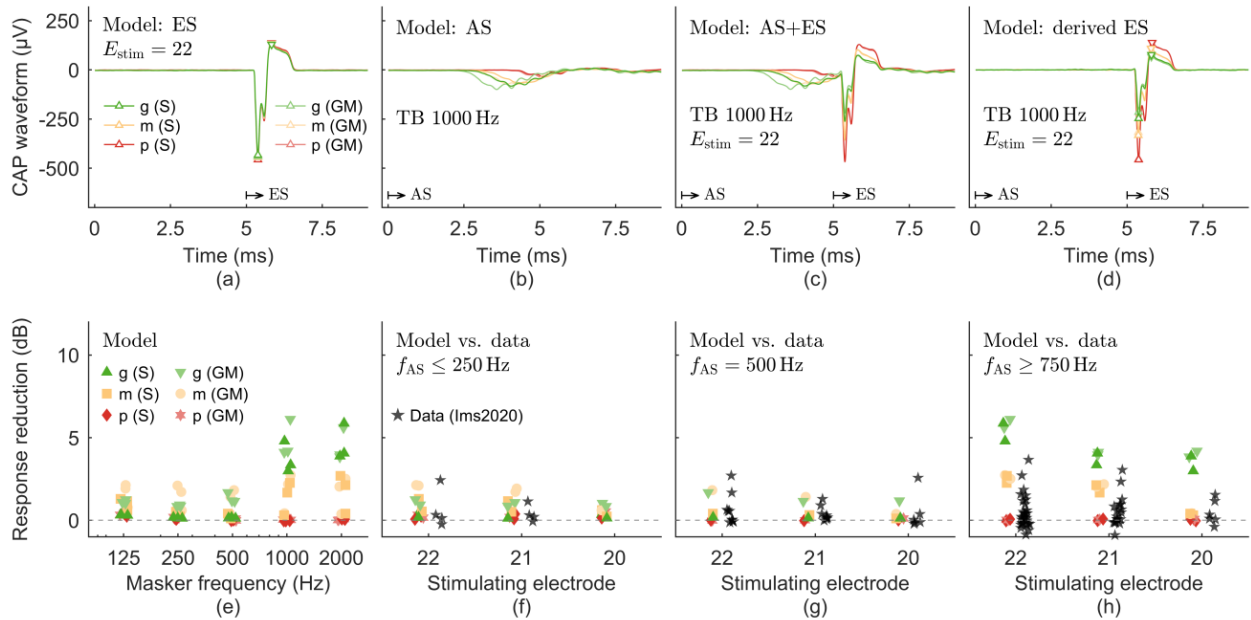

Fig. 5. Results for experiment 3 on acoustic masking of the electrically evoked compound action potential (CAP). 20 ms tone bursts as acoustic maskers and single electric pulses as probes were presented at M-level with a 5 ms masker-to-probe interval. **(a)–(d)** Example of simulated CAP waveforms for all hearing loss conditions, a 1000 Hz acoustic masker, electric stimulation on E22, and recording on E21. Small arrows at the bottom depict the stimulus onsets. **(a)** Unmasked responses to the electric probe. N1 and P1 peaks are denoted with triangles. **(b)** Responses to the acoustic masker. **(c)** Responses to masker and probe presented simultaneously. **(d)** Derived electric response: derived ES = (AS + ES) − AS. **(e)** Predicted response reduction of the derived electric response with respect to the electric-only response across different masker frequencies. Responses for stimulating electrodes 20, 21, and 22 were pooled together. **(f)–(h)** Comparison of predicted response reductions and human data of Imsiecke et al. ("Ims2020"; [13]) across masker frequencies and stimulating electrodes.

published data obtained from 13 Nuclus 24R(CS) and Nucleus 24RE(CA) users ("Hug2010"; [38]) as well as 4 Nucleus CI24M users ("Coh2004"; [37]). The predicted spatial spread patterns mirrored the human data, showing a decline in eCAP amplitude as the recording site moved further from the stimulation site. The spatial spread profile for the basal electrode partially deviated from this general pattern by including a second rise in the apical direction at distances greater than approximately 4.5 mm from the stimulation site (Fig. 3i). If unitary response models were used instead of the proposed SFCCs, the resulting spatial spread profiles would be constant (grey dotted lines in Fig. 3i,j).

### C. Experiment 2: aCAP AGFs

In experiment 2, aCAP responses to clicks, chirps, and 500 Hz tone bursts were investigated (Fig. 4). The aCAP waveforms predicted for clicks and chirps (Fig. 4a–b) showed clear N1 and P1 peaks that decreased in latency with increasing stimulation level. The waveforms for the 500 Hz tone burst also had clear N1 peaks, but less pronounced P1 peaks that were superimposed with an oscillating pattern at later latencies. These observations were consistent with results from human EAS subjects reported in [63]. When comparing the aCAP waveforms predicted at a fixed level of 75 dBnHL for different degrees of hearing loss (Fig. 4d–f), it was apparent that better residual hearing was associated with lower latencies and larger N1-P1 amplitudes for clicks and chirps, whereas the results for the 500 Hz tone burst were mixed.

Across low and medium stimulation levels (≤ 75 dBnHL), aCAP amplitudes increased as a function of level for all stimuli (Fig. 4g–i). This was in agreement with human aCAP data (compare Fig. 11A–C of [63]). However, there were at least two notable differences between the model predictions and the human data. Firstly, the predicted aCAP amplitudes were generally higher than those observed in human subjects by a factor of about three. Secondly, the predicted aCAP amplitudes for chirps began to decrease again when the stimulation level was further increased beyond 75 dBnHL. The reason was that asynchronous components were introduced in the PSTHs for chirps at high stimulation levels (not shown). The model correctly predicted that aCAP amplitudes for clicks and chirps were approximately two to three times larger than the responses to 500 Hz tone bursts.

### D. Experiment 3: acoustic masking of eCAPs

Experiment 3 investigated the interaction between simultaneous ES and AS with an acoustic-on-electric masking paradigm (Fig. 5). CAP responses were simulated for ES, AS, and combined AS+ES (Fig. 5a–c). A derived electric response was calculated by subtracting the response to the AS masker from the combined AS+ES response (Fig. 5d). The predicted N1-P1 amplitudes of the derived electric response were reduced with respect to the amplitudes of the ES-only response due to the influence of the acoustic masker.

Fig. 5e shows that this response reduction depended on both the frequency of the acoustic masker as well as on the degree of residual hearing. When the residual hearing was poor, the predicted responses were unaffected by the acoustic maskers and the response reduction was close to zero. Larger response reductions were predicted for medium and good residual hearing.

The predicted response reductions were similar to the reduction observed in human subjects by ("Ims2020", [13]; Fig. 5f–h). For maskers with frequencies of 500 Hz or below, the predicted response reductions were limited to 0–2 dB and were independent of the stimulating electrode (Fig. 5f,g). This was consistent with findings of [13], except for three experimental conditions that resulted in slightly higher response reductions of up to 2.7 dB. Larger response reductions, up to 6.0 dB (model) and 3.6 dB (data), were observed for maskers with higher frequencies (Fig. 5h). Here, both the predicted and human data indicated that stimulating more apical electrodes leads to a larger reduction of the eCAP response.

In addition to the results obtained when ES was presented at M-level, a series of simulations was performed where the electric probe level was set to 0% DR or 50% DR while keeping the acoustic maskers fixed at M-level. The results of these simulations are shown in the Supplementary Material (Suppl. Fig. 3). No response reduction was observed for poor residual hearing across all investigated probe levels. Similar to the results obtained at M-level, the response reductions for good and medium residual hearing were much smaller for masker frequencies of 500 Hz or below than those obtained at higher masker frequencies. Interestingly, lower probe levels led to a larger response reduction for good residual hearing (up to 13 dB at T-level for 1000 Hz and 2000 Hz maskers), whereas the response reduction for medium residual hearing decreased with decreasing probe levels.

## V. Discussion

A computational model was proposed to simulate the CAP of the auditory nerve in response to ES and AS in subjects with a CI and low-frequency residual hearing. The individual SFCCs predicted by the combined 3D FEM model and the multi-compartment neuron model strongly depended on the relative orientation of the recording electrode and the ANF, as well as on the location along the ANF were the excitation occurred. This was evident by distinct SFCCs that were obtained for ES when the ANF, stimulating electrode, or recording electrode was changed (Suppl. Fig. 2). Moreover, the individual SFCCs used in the present study resulted in a realistic spatial spread measured when the recording electrode was varied along the CI electrode array (Fig. 3i,j). These observations support the hypothesis that ANFs do not contribute uniformly to the CAP in the sense of the unitary response approach.

The morphology of the predicted eCAP and aCAP waveforms largely resembled those found in human recordings. They exhibited clear N1 and P1 peaks and distinct patterns for electric pulses and acoustic clicks, chirps, and tone bursts (Fig. 3a,b; Fig 4a–f). For ES, in some constellations of the stimulating and recording electrodes, a small second negative peak or plateau quickly following the N1 peak was visible in the simulated eCAP waveforms (Fig. 3b, Fig. 5a,c,d), which has not been reported in human recordings. This negative deflection was also observed in some individual SFCCs, usually in ANFs that were excited in their peripheral axon, and occurred around the time when the action potential passed the somatic region (Suppl. Fig. 1f; Suppl. Fig. 2a,b). The ANF morphology used in this study includes an additional pre-somatic region that was necessary to ensure spike conduction through the soma [42]. It is possible that this additional region contributed to the observed deflection. Using an updated multi-compartment model that is tailored more closely to the human ANF morphology may be beneficial in predicting the SFCCs [65]. The morphology of the aCAP waveforms in response to clicks, chirps, and tone bursts was very similar to the recordings of [63].

Comparing the characteristics of simulated eCAP AGFs to human data was challenging because these measures were likely influenced by differences in the recording protocols (Fig. 3d–g). Nevertheless, the predicted eCAP thresholds, dynamic ranges, linear slopes, and maximum amplitudes were within the range of the human data. Additionally, the model predicted aCAP AGFs in response to clicks, chirps, and 500 Hz tone bursts. The model predicted correctly that the amplitudes to 500 Hz tone bursts were significantly smaller than those in response to clicks and chirps. However, the predicted amplitudes to all acoustic stimuli were about three times larger than those recorded in humans [63]. One possible explanation for this discrepancy is the difficulty in calibrating stimulation levels for the model. The 1 mm criterion for estimating perceptual T-levels was a simple and practical solution that has been used in previous studies [42], [65], [66], but it may not always result in optimal estimations of the T-levels. Another possible reason is that synaptopathy and neural degeneration were only roughly considered in the model. All apical ANFs up to the characteristic frequency where the audiometric hearing loss reached 120 dB HL were simulated using the combined EAS spiking model and thus received input from an IHC (Fig. 2d). Histological studies suggest that aging and hearing loss are associated with the degeneration of ANFs, which may result in synaptopathy (i.e. disconnection from their IHC) or even loss of ANFs [67], [68]. Since fixed stimulation levels were used in experiment 2, synaptopathy and ANF loss could lead to a decrease in the number of ANFs that respond to the stimulus, resulting in a reduction of the predicted aCAP amplitudes. Lastly, the simulations may be constrained by the use of spiking models primarily derived from animal data [48], [51]. Since electrophysiological recordings from humans are unavailable for many stages of the auditory pathway, such as hair cells or single ANFs, the fitting of the two submodels for ES and AS heavily relies on data from animals, possibly resulting in disparities in ANF activity when compared to humans.

The model generally captured the electric-acoustic interaction investigated in experiment 3 well. AS maskers of varying frequencies significantly reduced the response amplitude to ES presented at M-level (Fig. 5). The model predicted that the response reduction (in dB) increases for more apical ES stimulation sites, as well as for higher AS masker frequencies up to 2000 Hz. This means that the response reduction was larger when the cochlea sites of acoustic and electric excitation were closer together, considering that the location of the most apical electrode corresponded with a tonotopic frequency of 1975 Hz. This pattern has been reported in previous electrophysiological [1], [69] and psychoacoustic [2], [4], [6], [7] masking experiments. However, it was less clear in the particular experiment of [13]. The additional simulations for ES probes presented at 0% DR or 50% DR indicated that it may be beneficial to explore electric-acoustic masking of CAPs at lower probe levels as well (Suppl. Fig. 3).

Similar to the 1 mm criterion for T-levels, the calibration of acoustic and electric M-levels poses a limitation of this study. While M-levels for ES were estimated aiming for a 4 mm spread of excitation along the OC as in prior modelling studies [32], [59], [65], [70], this same criterion proved inadequate for achieving realistic dynamic ranges for AS. More sophisticated methods need to be developed to accurately estimate perceptual loudness levels in computational models in future.

Instead of using separate models to predict PSTHs and SFCCs, it would be possible to generate CAPs for EAS in a unified manner by directly feeding the excitatory post-synaptic currents from the AS submodel into the multi-compartment SEF model. However, the setup with a fast phenomenological model for predicting PSTHs to EAS and the physiological SEF model for computing the SFCCs was preferred to reduce the computational costs. Simulations with the point-neuron submodel for ES were significantly faster than those with the multi-compartment SEF model, resulting in a 30-40 fold reduction in runtime for the simulations presented in this study.

## VI. CONCLUSION

A computational model was proposed to simulate the CAP of the auditory nerve in response to ES and AS in individuals with a CI and low-frequency residual hearing. The model was able to simulate realistic eCAP and aCAP responses, as well as CAP responses to combined EAS (acoustic-on-electric masking). Additionally, the model correctly simulated the spatial spread of the CAP response, which can be measured across different recording sites along the CI electrode array. The model is relevant for the design and testing of new electrophysiological measures. Specifically, the status of hair cells and auditory nerve fibers in human CI users may contribute to the large variability in CI implantation outcomes, but remains typically unknown. Ongoing research is focused on using CAPs to objectively estimate residual hearing, hair cell status, or neural survival. The proposed model contributes to this research by providing an efficient test bench for new experimental designs before moving to human subjects.

## ACKNOWLEDGMENT

The authors would like to thank Eugen Kludt and Nadine Buczak for help with the transimpedance measurements. The authors are also thankful to Jeong-Seo Kim for giving advice and supplying additional information on her studies.

# Supplementary Material for:

# A Computational Model of the Electrically or Acoustically Evoked Compound Action Potential in Cochlear Implant Users with Residual Hearing

Daniel Kipping[*], Yixuan Zhang, and Waldo Nogueira

## VII. Overview

This supplementary material contains additional information and simulation results.

Suppl. Fig. 1 illustrates the steps for the generation of single-fiber CAP contributions (SFCCs) for an auditory nerve fiber (ANF) that was close to the stimulating-recording electrode pair (left column) and another ANF located on the opposite site of the cochlea. The excitation of the ANF and propagation of an action potential (Suppl. Fig. 1b,h) was simulated with the nonlinear Schwarz & Eikhof-Frijns (SEF) multi-compartment model. The output was the trans-membrane current in each compartment over time (Suppl. Fig. 1c,i). These currents were used as current sources in the 3D FEM model to predict the time course of the voltage at the recording electrode. For each ANF, the simulation was performed in a supra-threshold condition where the stimulus level was just above the threshold to elicit an action potential (supra-threshold recording, Suppl. Fig. 1d,j), and in a sub-threshold condition where the stimulus level was just below threshold (sub-threshold recording, Suppl. Fig 1e,k). The sub-threshold recording was scaled to the supra-threshold amplitude to obtain a template of the stimulus artifact in the supra-threshold condition. This artifact template was subtracted from the supra-threshold recording to eliminate the stimulation artifact (equation 6 in the main manuscript), resulting in the raw SFCC (black line in Suppl. Fig. 1f,l). The raw SFCC still contained a second artifact that was produced at the time $t_{arrival}$ when the action potential reached the central end of the simulated ANF. This artifact was removed by fitting a cubic spline replacing the raw SFCC in the small time interval from 0.1 ms before to 0.15 ms after $t_{arrival}$. The final SFCC is shown with green lines in Suppl. Fig. 1f,l.

Suppl. Fig. 2 shows examples of SFCCs for ES for different ANFs and different constellations of the stimulating and recording electrodes. For a fixed pair of stimulating and recording electrodes, SFCCs differed significantly between different ANFs (Suppl. Fig. 2a). When the ANF and the recording electrode were fixed, SFCCs differed in a similar way depending on the recording electrode (Suppl. Fig. 2b). SFCCs depended also on the stimulation site when the ANF and recording electrode were fixed (Suppl. Fig. 2c).

Suppl. Fig. 3 presents additional results for experiment 3 on acoustic masking of the electrically evoked CAP. The main manuscript contains results for the acoustic masker and the electric probe presented at M-level (100% DR). Here, the level of the electric probe pulse was varied between 0% DR (T-level), 50% DR, and 100% DR (M-level). Rows correspond to the same stimulating and recording electrode pair, and columns correspond to the same frequency used for the acoustic tone burst masker. For masker frequencies of 125 Hz and 250 Hz, the amount of masking (response reduction of the eCAP in dB) was largely independent of probe level. For the 500 Hz masker, the response reduction for good and medium residual hearing changed with the probe level when the basilar membrane tuning according to Glasberg & Moore (GM) was used. A clear dependency for good and medium residual hearing can be seen for acoustic maskers presented at 1000 Hz or 2000 Hz and both versions of the basilar membrane tuning. The conditions with good or medium residual hearing resulted in opposite trends, where the response reduction for good residual hearing was largest at T-level whereas the response reduction for medium residual hearing was maximized at M-level.

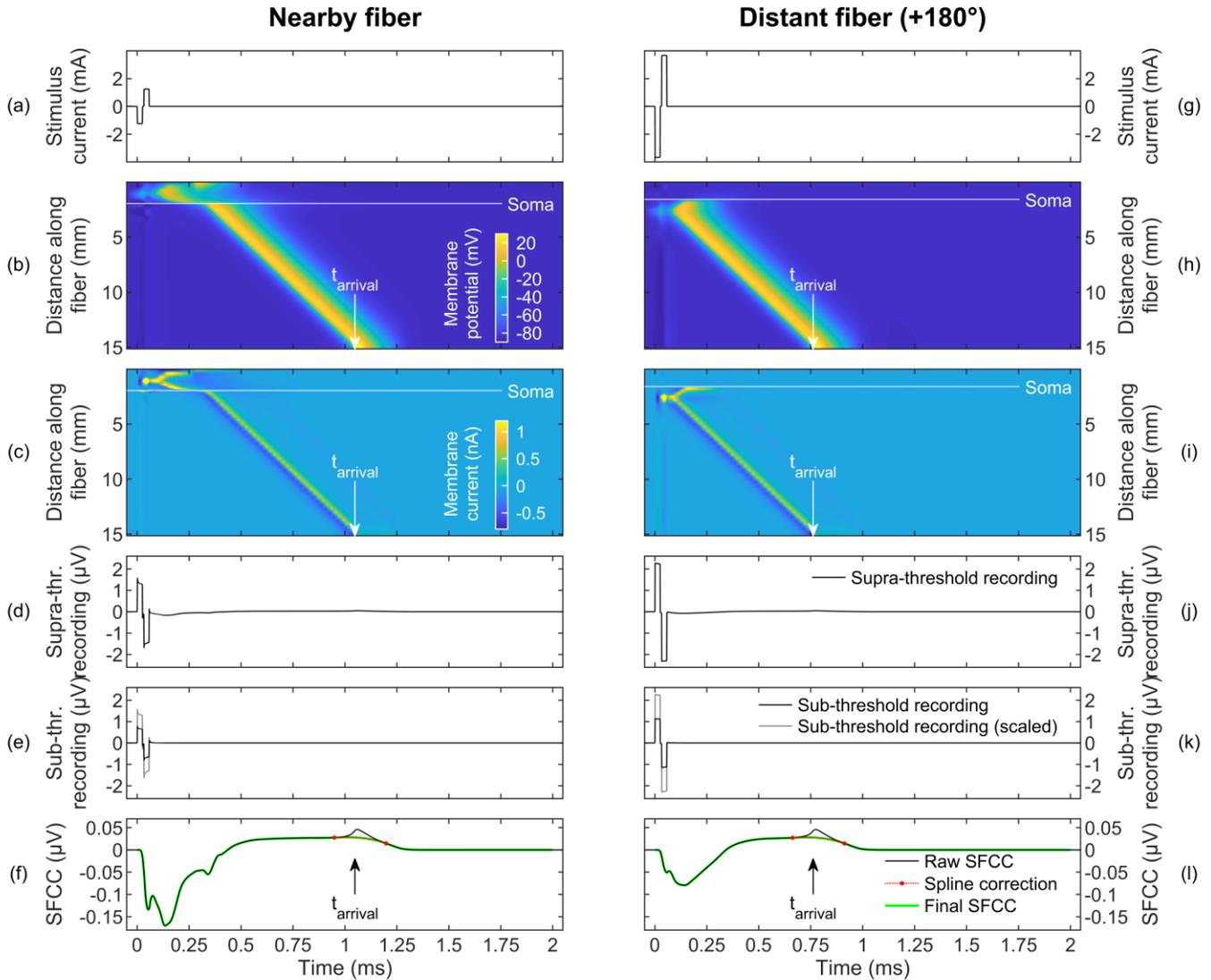

Suppl. Fig. 1. Generation of single-fiber CAP contributions (SFCCs) for electric stimulation on cochlear implant electrode E11 and recording on electrode E13. The left column (a)–(f) shows results for ANF no. 2600 which was located between E11 and E13, whereas the right column (g)–(l) shows corresponding results for ANF no. 1900 shifted by 180° in apical direction. **(a)** Biphasic current pulse used to activate the ANF. The stimulus level was adapted either to a supra-threshold level to evoke an action potential in the ANF, or to a sub-threshold level where no action potential was generated. **(b)** Cross-membrane potential of the ANF along time (x-axis) and length of the neuron (y-axis; 0 mm = peripheral terminal) for the supra-threshold condition. For this ANF, the action potential was generated in the peripheral axon (around 1 mm from the peripheral terminal) and propagated in central direction. The high capacitance of the somatic and pre-somatic compartments caused a delay of the action potential that was also described in Briaire and Frijns (2005). The arrival time $t_{arrival}$ of the action potential at the most central node of the model is marked with an arrow. **(c)** Cross-membrane current for the supra-threshold condition. Negative values correspond with an inward flow of positive charges (depolarization of the membrane), whereas positive values correspond with an outward flow of positive charges (repolarization of the membrane). **(d)** Recorded voltage at E13 for the supra-threshold condition. The neural response is dominated by the large stimulus artifact visible at the beginning of the recording. **(e)** Recorded voltage at E13 for the sub-threshold condition. The sub-threshold recording (black line), containing only the stimulus artifact without neural response, was scaled to match the stimulus artifact in the supra-threshold condition (gray line). **(f)** Raw SFCC (black line), generated by subtracting the scaled sub-threshold artifact template from the supra-threshold recording. On this scale, a second artifact is visible around $t_{arrival}$. This "border artifact" was produced when the propagation of the action potential stopped at the most central node. The border artifact was removed by fitting a spline to the time interval $[t_{arrival} - 0.1 \text{ ms}, t_{arrival} + 0.15 \text{ ms}]$ (red dotted line). The final SFCC was free of artifacts (green line). **(g)–(l)** Same as the left column, but for a distant ANF shifted apically by +180°. This ANF has a higher threshold level (g), and the action potential is elicited in the central axon (h),(i). The SFCC amplitude is lower than for the nearby ANF.

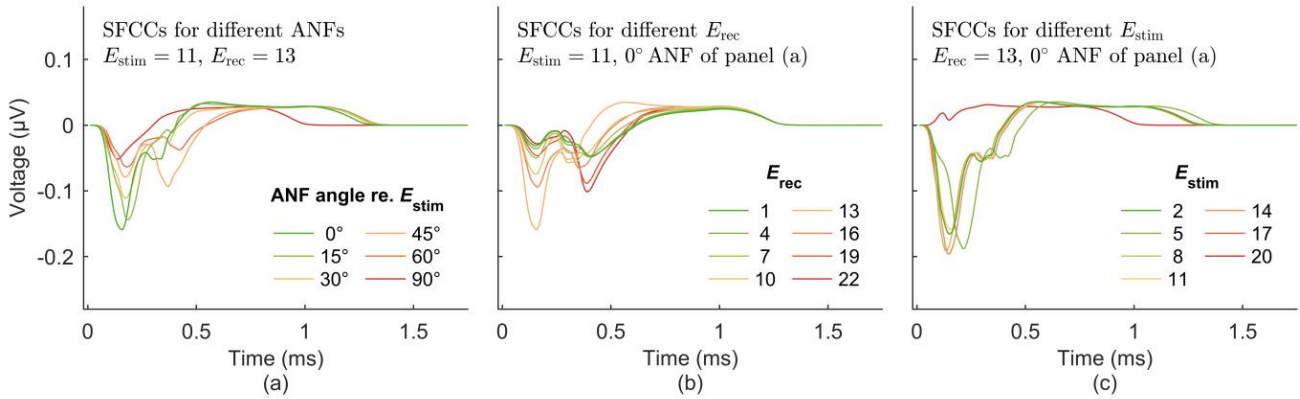

Suppl. Fig. 2. Examples of single-fiber CAP contributions (SFCCs) computed for electric stimulation. (a) SFCCs for different auditory nerve fibers (ANFs) when the stimulating and recording electrodes were fixed. The legend states the rotation angle (in apical direction) from the stimulating electrode to the ANF's peripheral terminal. (b) SFCCs for different recording electrodes when the ANF and the stimulating electrode were fixed. (c) SFCCs for different stimulating electrodes when the ANF and the recording electrode were fixed.

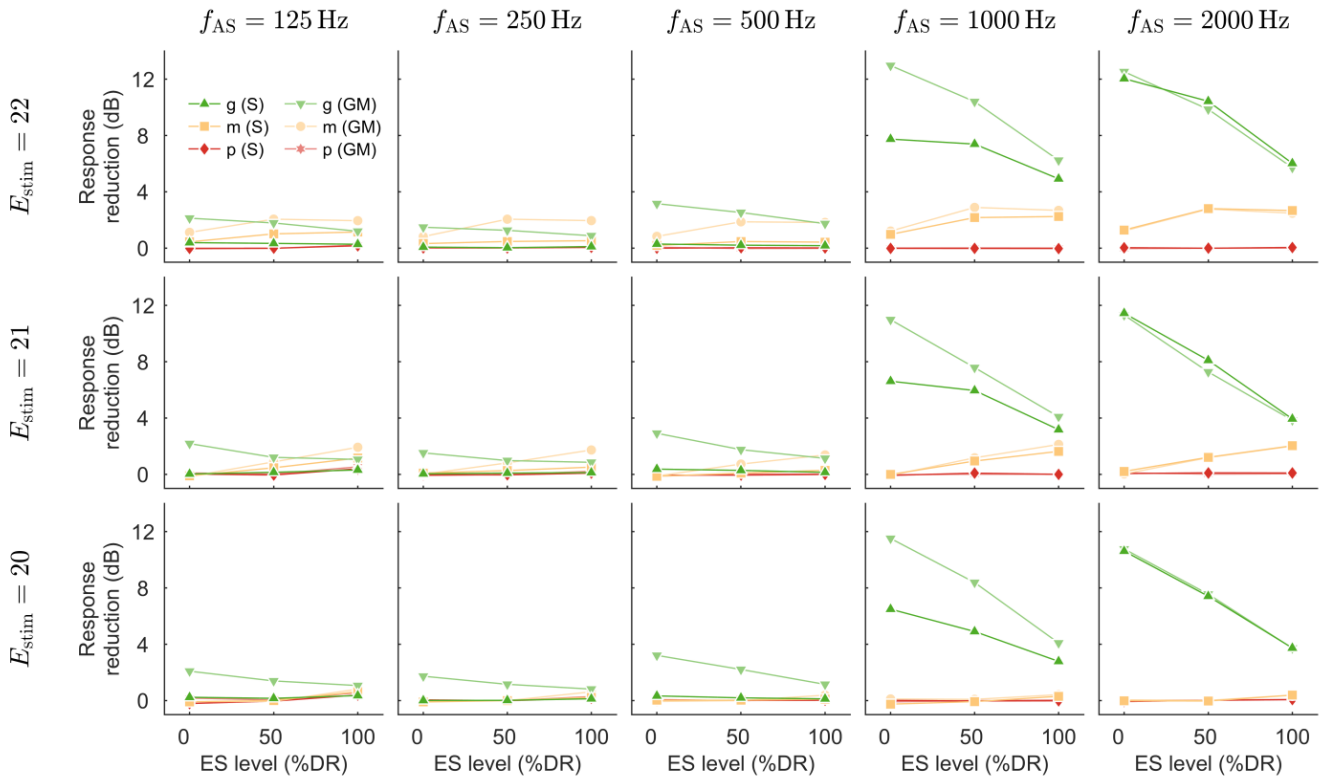

Suppl. Fig. 3. Additional results for experiment 3 on acoustic masking of the electrically evoked CAP for different electric probe levels. Each panel shows the response reduction of the derived electric response with respect to the electric only response for a specific pair of acoustic masker frequency ($f_{AS}$) and stimulating electrode ($E_{stim}$) across three electric probe levels: 0% DR (T-level), 50% DR, and 100% DR (M-level). Each row corresponds with the same stimulating electrode indicated on the left, and each column corresponds to the same masker frequency indicated on the top. Different colors and symbols depict the residual hearing (g – good; m – medium; p – poor) and the tuning of the basilar membrane filters (S – according to Shera et al.; GM – according to Glasberg and Moore).